\documentclass[
preprint,
superscriptaddress,
amsmath,amssymb,
aps,
pre,
floatfix,
]{revtex4-2}
\onecolumngrid
\usepackage[normalem]{ulem}
\usepackage{amsmath}
\usepackage{float}
\usepackage{amsfonts}
\usepackage{amssymb}
\usepackage{graphicx}
\usepackage{dcolumn}
\usepackage{bm}
\usepackage[breaklinks=true]{hyperref}
\usepackage{tikz}
\usetikzlibrary{arrows.meta} 
\usepackage{xcolor}
\usepackage{bm}
\newcommand{\hbf}[1]{\hat{\mathbf{#1}}}

\newcommand{\bsf}[1]{\textsf{\textbf{#1}}}

\begin{document}
\title{Edge states, pairing, and sorting  
of motile chiral particles} 
\author{Raushan Kant}
\email{raushankant@iisc.ac.in}
\affiliation{
Department of Physics, Indian Institute of Science, Bangalore 560 012, India}
\author{Ananyo Maitra}
\email{nyomaitra07@gmail.com}
\affiliation{Laboratoire de Physique Th\'{e}orique et Mod\'{e}lisation, CNRS UMR 8089, CY Cergy Paris Universit\'{e}, F-95032 Cergy-Pontoise Cedex, France}
\affiliation{Laboratoire Jean Perrin, UMR 8237 CNRS, Sorbonne Universit\'{e}, 75005 Paris, France}

\author{A K Sood}
\email{asood@iisc.ac.in}
\affiliation{Department of Physics, Indian Institute of Science, Bangalore 560 012, India}
\author{Sriram Ramaswamy}
\email{sriram@iisc.ac.in}
\affiliation{
Department of Physics, Indian Institute of Science, Bangalore 560 012, India}


\begin{abstract} 

We present experiments on chiral active polar particles, realised as vibrated granular rods, revealing 
evidence for ``skipping orbits'' at hard boundaries. These classical edge states exhibit a net circulation opposite to the particles' intrinsic rotation and lead to a pronounced accumulation at the boundary, stronger than for their achiral counterparts. The directed nature of these orbits provides a simple yet high-fidelity mechanism for chiral sorting -- even for solitary particles, unlike in T. Barois et al., \href{https://doi.org/10.1103/PhysRevLett.125.238003}{Phys. Rev. Lett. \textbf{125}, 238003 (2020)}. We propose a unified theoretical framework for boundary interactions of both chiral and achiral particles. In this model, an effective outward radial force, proportional to motility and chirality, explains the observed boundary-hugging. Our theory predicts, and our experiments confirm, a transition in the pairing of two particles of the same chirality, from apolar spinners to polar circle walkers, with increasing packing fraction of an ambient medium of beads.





\end{abstract}

\maketitle

 \section{Introduction} \label{intro}
\begin{figure}[h]
    \centering
    \includegraphics[width=0.99\linewidth]{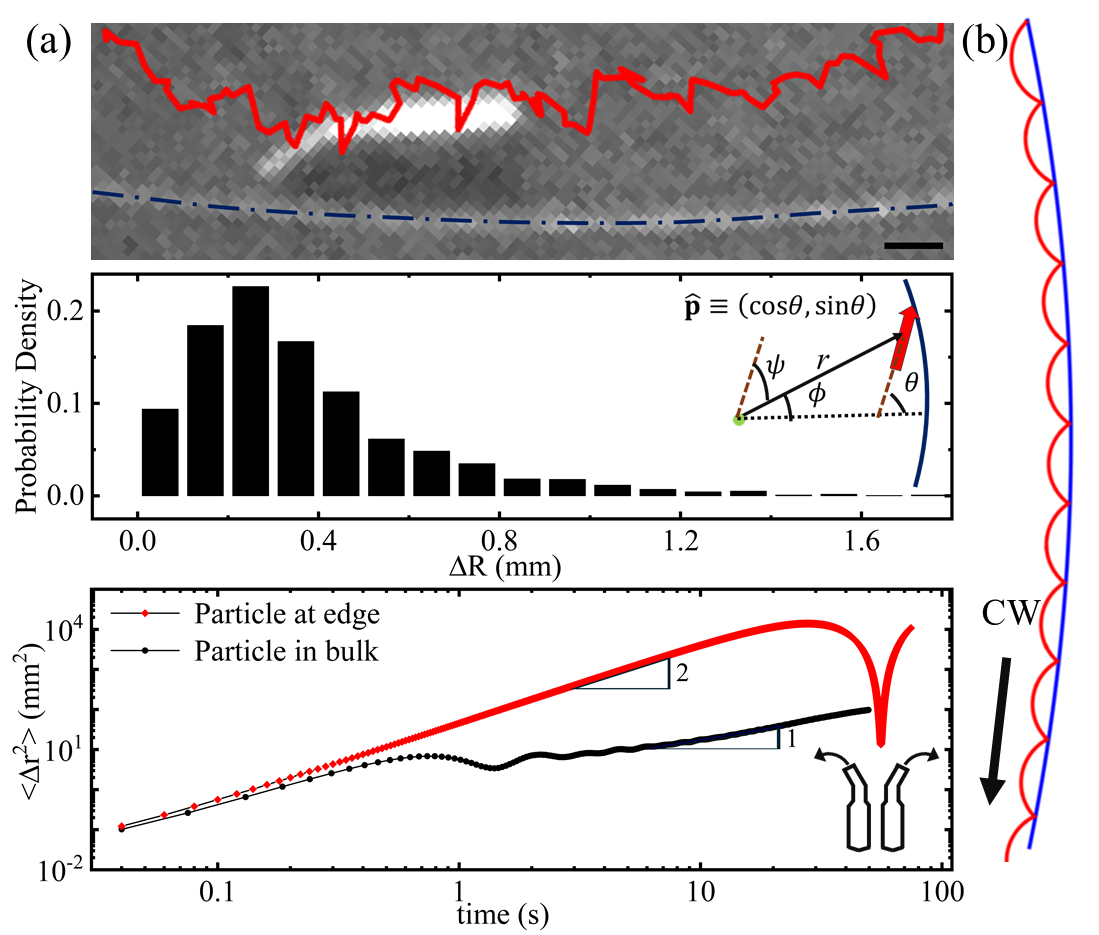}
    \caption{Chiral Active Polar Particles (CAPPs) at a boundary: (a) (Top) Noisy skipping orbit of CAPP along hard boundary. Overlaid red line tracks trajectory of particle's thick end over 2.1 seconds. Scale bar: 1 cm.
     (Middle) Histogram of distance from particle's thin end to wall, showing strong localisation. Inset: coordinates used in theoretical analysis---orientation \(\mathbf{\hat{p}}\equiv(\cos\theta,\sin\theta)\) (red arrow) and position $(r,\phi)$ of particle. \(\psi = \theta - \phi\) is orientation relative to polar angle. (Bottom) Mean Square Displacement analysis: CAPPs at edge (red) show persistent ballistic motion for approximately 10 seconds, with dip at longer times due to finite system size. Conversely, blue curves, representing average MSD of 14 particles in bulk, indicate diffusive motion over longer times. (Inset: Top right corner: outline drawing of 
    $\boldsymbol{\circlearrowleft}$ and $\boldsymbol{\circlearrowright}$ rotating CAPPs).  (b) Section of skipping trajectory (red) obtained by solving \eqref{rdot} and \eqref{psidot} with $\alpha=1$ and $\beta=10^6$ (i.e., $\zeta=10^6$) with a ``wall potential'' $U=e^{-10^4(\rho-\rho_0)}/(\rho-\rho_0)^{12}$, where $\rho_0$ is radial position of wall in reduced coordinates. Potential chosen to mimic hard wall, but qualitative shape of trajectory is insensitive to this detail, requiring only sharp decay away from $\rho_0$.}
    \label{OnlyChiral}
\end{figure}

\begin{figure}
    \centering
    \includegraphics[width=0.95\linewidth]{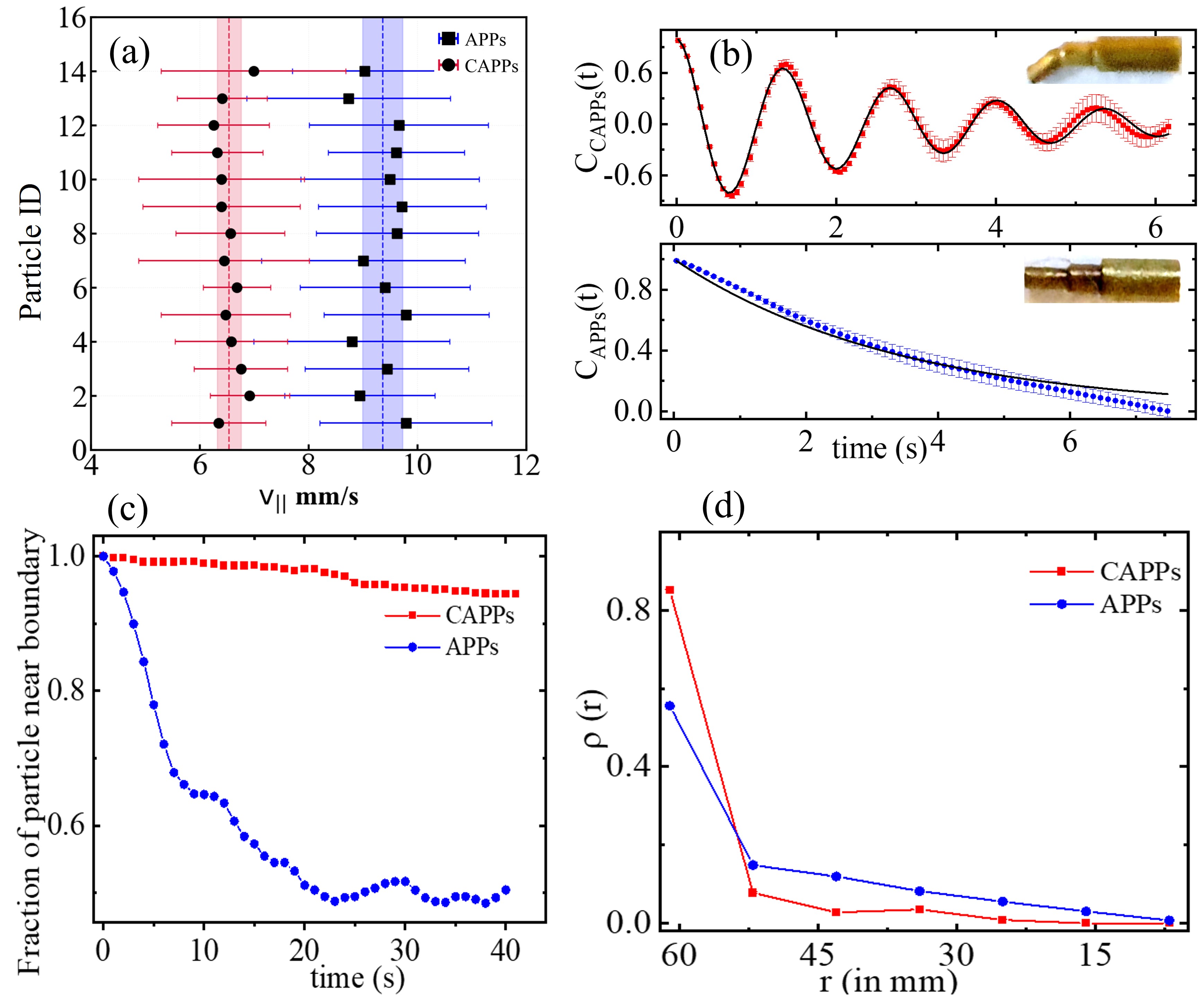}
    \caption{Self-propulsion speed \(V_{\parallel}\) averaged over 14 particles each for CAPPs (circles) and APPs (squares). (b) Top panel: fit (solid curve) of rotational autocorrelation of CAPPs (red squares) yields \(\tau_{c} = 3.1 \pm 0.15\) s and \(\Omega = 4.4 \pm 0.1\) rad/s. Note zero-crossing well before \(\tau_c\), at \(t = \frac{\pi}{2|\Omega|}\).    
    Bottom panel: rotational autocorrelation of APPs (blue circles); fit (solid line) yields \(\tau_{s} = 3.4 \pm 0.13\) s. 
    (c) Comparison of escape times for CAPPs (red squares) and APPs (blue circles) from edge to the bulk; particles of each type are initially positioned along the boundary (movie SM3 \cite{moviechiralch52025}). (d) Steady state distribution of radial positions of CAPPs (red squares) and APPs (blue circles); total number of particles \((N = 36)\). 
    }
    \label{ComparePolar}
\end{figure}

The structural asymmetries of an active
\cite{ReviewMarchettiSriram2013, ReviewSriram2010, Dadhichi_Jstat} object govern the nature of its self-driven motion, and chirality \cite{Lubensky_Kamien} is a ubiquitous asymmetry of the living world \cite{Ano_rev}. 
Chiral objects without polarity, when energized, rotate 
\cite{van2016spatiotemporal,scholz2018rotating,Workamp2018ymmetry,soni2019odd,Cid2021Arrested} or display ``odd diffusion'' \cite{chun2018emergence,hargus2021odd,banerjee2024active,kalz2022collisions,muzzeddu2025self,kalz2024oscillatory,langer2024dance,kalz2025reversal}. Chiral active polar particles \cite{liebchen2022chiral,Review2016BechingerActiveComplex,mecke2024emergent,Riedel2005Sperm,patra2022collective,diluzio2005escherichia,Leonardo2011Swimming,Lauga2006Swimming,alvarez2021reconfigurable,kummel2013circular,zhang2020reconfigurable,barois2020,arora2021emergent,chan2024chiral} (CAPPs), see Fig. \ref{OnlyChiral}(a) describe circular trajectories with fixed sense in the absence of noise and away from boundaries or other particles, much like 
charges in an external magnetic field.  
In this Letter, we present experiments and a theoretical understanding thereof on CAPPs 
in the form of bent polar particles, made active by vertical vibration of their support base and viewed through a transparent lid that effectively confines them to two dimensions and prevents them from flipping to invert their chirality.
We study single-particle and pair behaviour \cite{gupta2022active,arora2021emergent, Cid2021Arrested, Fily2012rotors} in a circular domain with a hard boundary, on a bare substrate and in a dense background of non-motile beads. We show that the CAPPs persistently move along the boundary roughly executing skipping orbits whose net circulation is opposite to that of their motion in the bulk. However, the orbital moment, far from cancelling as 
for charged Brownian particles at thermal equilibrium in a magnetic field, \cite{vanleeuwen1921electronic,bohr1972dissertation}, is \textit{dominated} by the edge: the CAPPs accumulate strongly at the boundary [Fig. \ref{OnlyChiral}(a) \(\&\) movie SM1 \cite{moviechiralch52025}], an aspect not emphasised in  \cite{Teeffelen2009Clockwise}. Chirality and polarity are crucial for boundary-hugging; the boundary-bulk occupancy contrast is modest for achiral Active Polar Particles (APPs) [Fig. \ref{ComparePolar}(d)]. 

\begin{figure}
    \centering
    \includegraphics[width=0.99\linewidth]{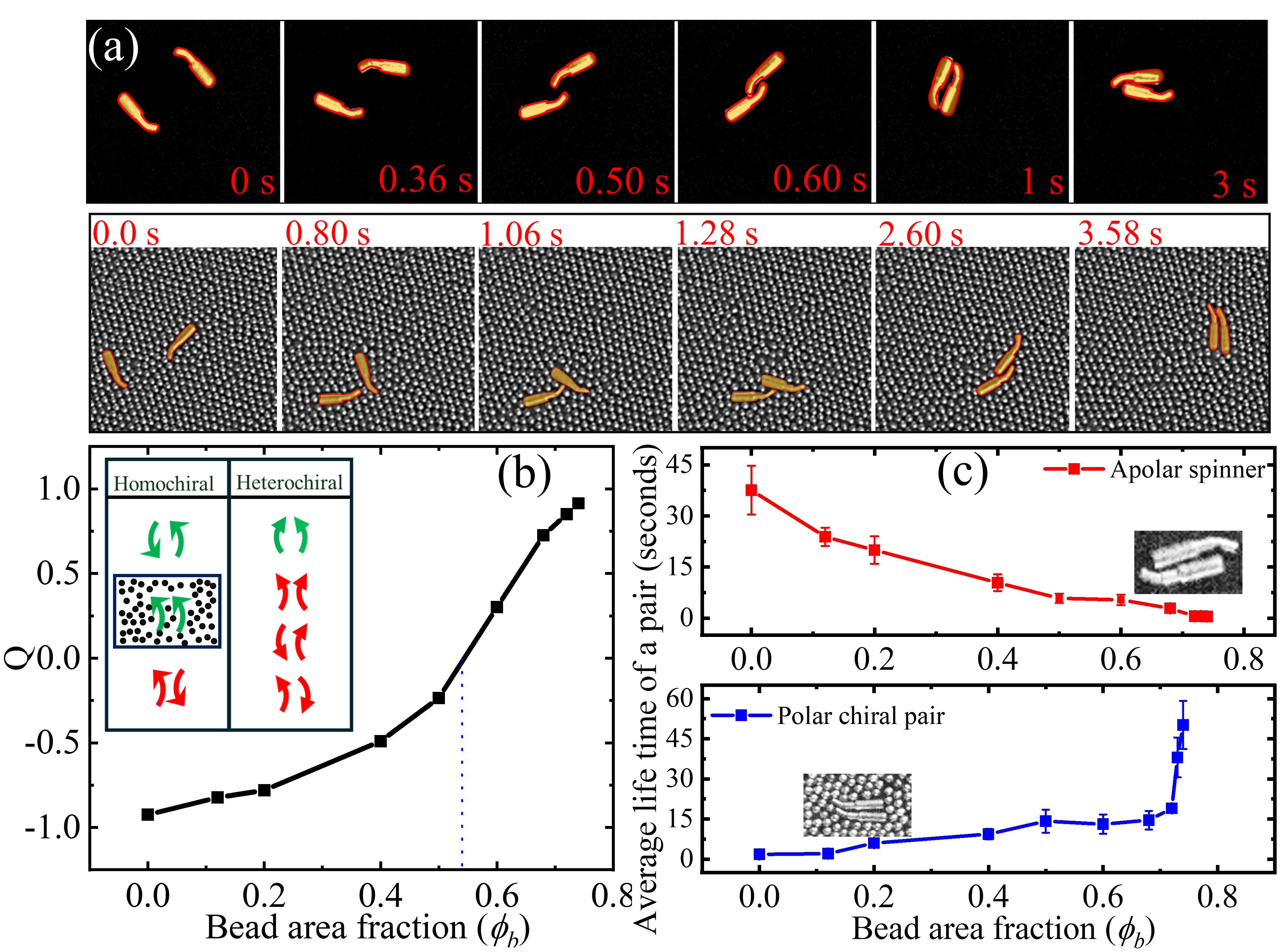}
    \caption{ (a) Top: time sequence of apolar chiral spinner (spinner) without bead medium. Bottom: polar chiral circle walker at \(\phi_{b} = 0.74\). (b) The pairing measure \(Q = \langle \cos(\Delta\theta) \rangle\) versus \(\phi_{b}\) confirms a shift in pairing preference from polar pair with increasing \(\phi_{b}\); (Inset: Two-particle pairing in chiral active particles (CAPPs). Viable (green) and unviable (red) pairings are shown for both homo- and hetero-type CAPPs, where dots in black are bead background). Blue dotted line indicates zero crossing of \(Q\). (c) Lifetimes of polar pair (top) and spinner (bottom) with increasing \(\phi_b\).}
    \label{Pairing}
\end{figure}

We construct a minimal single-particle theory in which the edge currents emerge naturally 
as directed skipping trajectories whose sense depends on the chirality. The theory predicts, and our experiments confirm (see Fig. \ref{Pairing}), apolar pairing 
of 
homochiral particles 
on a bare substrate, with a transition to polar alignment \cite{kumar2014flocking} with increasing concentration of a background bead medium, and polar alignment \cite{arora2021emergent} of a heterochiral pair. We also exploit the directed character of skipping orbits to chirally sort a mixture of CAPPs Fig. \ref{sorting}. Our process works even when only one particle is present at a time, unlike in \cite{barois2020} where interparticle collisions are essential to initiate sorting.

\section{Experimental details} 
\label{Exp_detail}
The achiral APPs are $4.5$ mm-long brass rods, tapered along their length from $1.1$ mm to $0.7$ mm. A bend at the narrow end turns them into CAPPs, [Fig. \ref{ComparePolar}(b), inset]. We confine the particles between a circular plate, with a diameter of \(12.2\) cm, and a glass lid separated by \(1.12\) mm. 
Vertical agitation of the particle in confined geometry generates both active force \cite{yamada2003asymmetric,Dorbolo2005Dimer} and torque. For details of our experimental setup, see \cite{Vijay2007LongGiant,kumar2014flocking,gupta2022active,kant2024Bulk}. 
Our guiding geometry for sorting a heterochiral mixture of CAPPs is engineered by glueing copper wires to form a cup-and-straw arrangement; [Fig. \ref{sorting}(a)].
We capture images at 25 FPS on a Redlake MotionPro X3 camera. For some of the sorting experiments, we used a Basler camera. Images were analysed using Fiji (ImageJ), MATLAB, and Python.

We denote a particle's chirality by $\boldsymbol{\circlearrowleft}$ or $\boldsymbol{\circlearrowright}$ depending on whether it rotates counter-clockwise or clockwise in the bulk (i.e., away from any boundaries); see movie SM2 \cite{moviechiralch52025}.
We extract the angular speed (\(\Omega\)) from the autocorrelation function for 14 chiral particles [Fig. \ref{ComparePolar}(b)]. In bulk, CAPPs rotate at an average angular speed of \(4.4 \pm 0.1\) rad/s. We calculate the velocity component \(V_{||}\) along the polarity direction for each of the 14 CAPPs by multiplying their angular speed by the radius of their circular trajectory, for at least 15 such trajectories [Fig. \ref{ComparePolar}(a)]. For APPs, we calculate \(V_{||}\) for 14 APPs with a frame gap proportional to one third of the persistence time (\(\tau_s\)) [Fig. \ref{ComparePolar}(a)]. Average \(V_{||}\) for CAPPs and APPs is \(6.5 \pm 0.05\)  and \(9.3 \pm 0.1\) mm/s respectively. As a result of the periodic rotation of the motility direction [Fig. \ref{OnlyChiral}(a)], the mean square displacement (MSD) of CAPPs is oscillatory in the bulk, with its first local minimum at time \(\approx 1.4\) s$=2\pi/|\Omega|$.
A \( \boldsymbol{\circlearrowleft}\)/\( \boldsymbol{\circlearrowright}\) particle can reach the boundary through diffusion and persistent motion and, localised to the edge,  traces a 
macroscopically CW/CCW (clockwise/counter-clockwise) orbit; i.e., there is a micro-macro chirality reversal. 
To account for the polar orientation of our particles, we use the unit vector \(\mathbf{\hat{p}} \equiv (\cos \theta, \sin \theta)\), which points towards the tapered end. 
A CAPP is characterised by the time $\pi/2|\Omega|$ it takes to rotate by an angle $\pi/2$. For APPs the timescale of relevance is the persistence time $\tau_{s} = 3.4 \pm 0.1 s$ of its orientation in the face of noise. Since $\tau_s\gg \pi/2|\Omega|$, one might naively have expected the residence time of APPs \cite{Caprini2019ChiralConfinement} at a hard wall to be much larger than that of CAPPs. 
We find instead that CAPPs are much more localised at the boundary than APPs.


We next present measurements quantifying the longer residence time, and hence higher density, of CAPPs at the boundary. 
(i) In separate experiments, we decorate the boundary with a monolayer of $72$ CAPPs and APPs to compare the time taken by each type to escape from the edge (distance from the particle centre to the circular boundary $< 1.5 l$) to the bulk. After $40$ s---the mean time it takes for half the APPs to escape to the bulk---more than $90$\% of CAPPs remain at the boundary [Fig. \ref{ComparePolar}(c) \(\&\) movie SM3 \cite{moviechiralch52025}]. (ii) We disperse \(N = 36\) particles in the bulk and measure the probability of finding them at the edge in the steady state. For this configuration, approximately \(85\%\) of the CAPPs are found at the edge, compared to \(55\%\) of the APPs [see Fig. \ref{ComparePolar}(d) and movie SM4 \cite{moviechiralch52025}]. For larger \(N\), intra-CAPP interactions lead to the formation of multimers and small rotating clusters in the bulk (see movie SM5 \cite{moviechiralch52025}), which dominate over particle-wall interactions.
The fraction of APPs at the wall is less affected by \(N\) than CAPPs, and their clustering near the wall at high \(N\) is similar to \cite{wensink2008aggregationSPPs}. 

\section{Theory} \label{theorysec}
A minimal model (similar to the one in \cite{carrillo2025depinning}; also see App \ref{min_theory} for a discussion of other contributions that may arise in a faithful dynamical description of CAPPs) generates the key features of the interaction of a CAPP with the boundary. 
The particle's centroid position and orientation are denoted by ${\bf r}$ and $\hbf{p}$ respectively. The particle has a self-propulsion speed $v_0$ along $\hbf{p}$, and its inertialess and deterministic equation of motion (see App. \ref{physgamma} for a discussion of the stochastic equations) reads 
\begin{equation}
\label{rEq}
\dot{\mathbf{r}} = v_0 \hbf{p} + \mu \mathbf{F}({\bf r}) 
\,. 
\end{equation}
Here ${\bf F}$ is the force exerted on the particle by the wall or other particles and $\mu$ is the mobility. 
The orientational dynamics is 
\begin{equation}
\label{pEq}
 \dot {\hbf{p}} = \gamma \bm{\Pi}\cdot \mathbf{F} ({\bf r}) {- \Omega\bm{\epsilon}}\cdot\hbf{p}\,,
 \end{equation}
where $\bm{\epsilon}$ is the two-dimensional Levi-Civita tensor, and $\bm{\Pi} = \bsf{I} - \hbf{p} \hbf{p}$, so that the $\gamma$ term \cite{Sano_birds,Brotto,kumar2014flocking,self_align_rev} acts as long as $\hbf{p}$ is not parallel to ${\bf F}$. 
$\Omega$, the constant active rate of 
rotation of the direction of $\hbf{p}$ \cite{Ano_rev, Ano_hex} in the absence of ${\bf F}$, is the only effect of chirality that we consider.
\(\Omega > 0\) and $<0$ refer respectively to \(\circlearrowleft\) and \(\circlearrowright\) CAPPs.

$\mu$ and $\gamma$ control the interaction of the particle with the wall, $\mu$ simply pushing it away and $\gamma>0 \, (<0)$ tending to turn its polarity towards (away from) ${\bf F}$ if it is not perfectly normal to the wall. In our experiments $\gamma>0$. A force ${\bf F}$ that acts for a short time and is large (i.e., is derived from a sharp near-step-function potential) leads, in conjunction with self-propulsion, to an active particle being reflected at a flat wall (except when normally incident) even when $\mu=0$. In this sense, $\hat{{\bf p}}$ in \eqref{rEq} and \eqref{pEq} acts like an ersatz momentum corresponding to ${\bf r}$ at least in the limit $\mu=0$. In this case, the chiral term $\Omega$ can be viewed either as a chiral velocity-dependent force---an odd friction \cite{Ano_hex, Ano_rev}---or as a Lorentz force due to a constant magnetic field normal to the plate.

The model for edge currents of CAPPs at walls immediately presents itself, with direction of propagation determined by chirality. A $\boldsymbol{\circlearrowleft} (\boldsymbol{\circlearrowright})$ particle moves into the wall on its left (right). Upon encountering the wall, it is ``reflected'' back, i.e., its polarity direction reverses because of $\gamma$, but because of its chirality, it again runs into the wall, thus performing a CW (CCW) orbit at a circular boundary or, equivalently, moving up (down) a straight one.

We now present a more detailed analysis of Eqs. \eqref{rEq} and \eqref{pEq} assuming a purely radial boundary, implying ${\bf F}\equiv F(r)\hbf{r}$, and rewrite \eqref{rEq} and \eqref{pEq} in polar coordinates $(r,\phi)$; the conceptually similar case of a straight boundary is discussed in App. \ref{EOM_straight}. We use non-dimensional variables $\tau=t|\Omega|$, $\rho=r|\Omega|/v_0$ (i.e., distance and time are non-dimensionalised, respectively, by the radius and the time-period of rotation of the CAPP away from the boundaries) and parameters $\alpha=\mu/v_0$ and $\beta=\gamma/|\Omega|$.
The deterministic dynamics reduces to that of two variables $r$ and the angle $\psi\equiv\theta-\phi$, see Fig. \ref{OnlyChiral}(a), that $\hbf{p}$ makes with the radial direction:  
\begin{equation}
\label{rdot}
    \frac{d\rho}{d\tau}=\cos\psi+\alpha F(\rho)
    \,,
\end{equation}
\begin{equation}
\label{psidot}
   \frac{d\psi}{d\tau}={\omega} -\left[\beta F(\rho)+\frac{1}{\rho}\right]\sin\psi
    \,.
\end{equation}
where $\omega=\pm 1$.


We model the wall by a potential that rises steeply at a particular $\rho$, but is not a step function. In that case, the opposing force due to the wall balances the radial component of self-propulsion at some $\rho = \rho_0$. Eqs. \eqref{rdot} and \eqref{psidot} then have a steady-state solution $(\rho_0,\psi_0)$,
\begin{equation}
    \omega=\sin\psi_0\left(\frac{1}{\rho_0}-\frac{\beta}{\alpha}\cos\psi_0\right),\,\,\, \alpha F(\rho_0)=-\cos\psi_0\,.
    \label{omgcond}
\end{equation}
The first condition is satisfied provided $|1/\rho_0-\zeta|$ is large enough, where the dimensionless parameter $\zeta\equiv\beta/\alpha=\gamma v_0/\mu|\Omega|$.
For $\omega=+1$, Eq. \eqref{omgcond} has a pair of CW solutions with $0 > \psi_0 > -\cos^{-1} 1/\zeta\rho_0$ in the relevant domain $[-\pi/2,\pi/2]$---of which the more negative solution is the one that is stable for large $\rho_0$ and $\zeta$, as we show in App. \ref{lin_stab}---which means the walker is turned \textit{towards} the wall. As $\zeta\to\infty$, i.e., when $\gamma/\mu\to\infty$, these solutions go to $\psi_0\to 0$ and $\psi_0\to-\pi/2$.
Via \eqref{rdot}, this implies an effective force towards positive $\rho$, pushing the particle into the wall, whose hard repulsion combines to create a confining well. This \emph{deterministic} mechanism for density enhancement at the wall is distinct from the stochastic one in achiral active Brownian particles \cite{Duzgun} with $\gamma=0$.
As the dynamics in the experiment is noisy, we expect the particle to be found within a typical small distance from the wall as is indeed the case (see Fig. \ref{OnlyChiral}a). 
Still for $\omega=+1$, a CCW trajectory close to the wall cannot survive as the walker would turn away from the wall. The above arguments, \textit{mutatis mutandis}, can be applied to show that $\omega=-1$ yields wall-hugging CCW trajectories.



Unfortunately, it is difficult to measure $\zeta$ directly in our experiments. In App. \ref{zeta_measure}, we discuss a method of estimating $\mu/\gamma$ from the steady-state angle that an \emph{APP}---which does not display a skipping orbit---makes with the wall. Using the $v_0$ and $\Omega$ we measure for our CAPP we find that $\zeta\to\infty$ (i.e., very large) in our experiments, assuming the same $\mu/\gamma$ for CAPPs as well. This assumption is not critical, for, as we remark below, we find skipping orbits for a wide range of parameter values.

A numerical solution of \eqref{rdot} and \eqref{psidot} with a steep potential and $\beta\gg \alpha$, i.e., for very large $\zeta$, displays a skipping orbit (see Fig. \ref{OnlyChiral}b):
even without noise, the particle doesn't simply settle down to the steady state discussed above, as it does at a somewhat lower $\zeta$; instead, because of the combined effect of the inertia-like effect of activity and chirality-induced turning, it ``bounces'' along the wall. This is best seen in the $\alpha=0$ ($\zeta\to\infty$) limit, in which the skipping orbits survive, see Fig. \ref{nomob} in the appendices, and the $\rho-\psi$ dynamics 
to linear order in $\chi=\pi/2-\psi$ reads
\begin{equation}
    \frac{d\rho}{d\tau}=\chi,\,\,\,\frac{d\chi}{d\tau}=\frac{1+\omega\rho}{\rho}+\beta F(\rho)\,,
\end{equation}
with $\rho$ and $\chi$ acting like position and momentum. Mechanically, for a localised $F(\rho_0)$, the force acts only when the particle is at $\rho_0$ and reverses its ``momentum'' $\chi$. The particle then moves away from the wall and enters a force-free region. The free chiral rotation of the particle (like gravitational free-fall in this analogy) then again takes it towards the wall, thus yielding the skipping orbits. That such skipping orbits are realised by our particle is suggested by the trajectory displayed in Fig. \ref{OnlyChiral}a and movie SM1 \cite{moviechiralch52025}, though the data is too noisy to measure a characteristic frequency. 


Our minimal theory shows that edge states induced by skipping orbits should generically emerge in a wide range of CAPPs, irrespective of manufacturing details. This significantly expands upon the work of \cite{Teeffelen2009Clockwise}, which discusses a rich range of behaviours of CAPPs confined to a disc or an annulus, but does not emphasise the role of self-alignment or boundary-hugging orbits. 

\section{Implications for pair interactions and chiral sorting} \label{implications}
The pair interactions of CAPPs provide a test of our theory. {We have seen above} that a single $ \boldsymbol{\circlearrowleft}$ ($ \boldsymbol{\circlearrowright}$) CAPP moves preferentially with the bounding wall on the left (right), i.e. they align their polarity with the direction in which they would roll against the wall if they were pure spinners. We know in addition from \cite{kumar2014flocking} that polar particles align with each other through the flows they generate by moving through a background of beads. We show now that these independent processes \textit{predict} the nature of interactions between pairs of CAPPs as a function of their relative chirality, on a bare substrate and in a bead medium. Consider first two CAPPs of the same chirality, say $\boldsymbol{\circlearrowleft}$, in the absence of a medium. For each CAPP, the other acts as a steric ``wall'' and for both of them to have the other to their left, they must form an \textit{apolar} pair, that is, a non-motile \textit{spinner} \cite{Cid2021Arrested, Fily2012rotors,tan2022odd,kalz2022collisions,petroff2015fast, Dabra2025Depletion} [Fig. \ref{Pairing}(a)]. If they formed a \textit{polar} pair, i.e., a circle walker like the individual constituents, as in Fig. \ref{Pairing}(a), the outer CAPP would be to the right of the inner CAPP. For a heterochiral pair, on the other hand, the same argument predicts \textit{polar} alignment in the form of a \textit{mover} \cite{arora2021emergent} [Inset, Fig. \ref{Pairing}(b), movie SM8 \cite{moviechiralch52025}]. A dense non-motile background of beads, if present, will be set in motion by frictional and steric coupling to the motile polar rods \cite{kumar2014flocking}. The polarisation of other particles responds to this bead velocity field via a coupling $\dot{{\bf p}}_i\propto {\bf p}_i\times [{\bf p}_i\times{\bf v}({\bf r}_i)]$, where ${\bf p}_i$ is the polarisation of the $i-$th particle that is being reoriented by the beads and ${\bf v}({\bf r}_i)$ is the value of the joint velocity field (of the particles and the beads) evaluated at the site of the particle. Therefore, the bead medium fundamentally modifies the interaction of the CAPPs: they no longer interact directly but via this flow-induced alignment interaction \cite{kumar2014flocking}, which is related to the $\gamma$ term in \eqref{pEq} (see App. \ref{physgamma}). In contrast to the direct interaction, this mechanism always tends to align the overall polarity of CAPPs with each other.
For a homochiral pair, this mechanism shifts the balance away from non-motile spinners, in favour of polar circle walkers. The already polar pairing of heterochiral dimers should remain unaffected. 

Our experimental study fully confirms these predictions, establishing a switch in the dominant type of homochiral pairing, from apolar spinners to polar circle walkers, with increasing \(\phi_{b}\) [Fig. \ref{Pairing}(b), movie SM9 \cite{moviechiralch52025}].
We adopt the definition that two particles are declared a pair if the distance between their centroids is less than \(2 l\). To quantify the transition in the nature of pairing of CAPPs as a function of the concentration of the bead medium, we define a \textit{pairing preference} $Q = \langle \cos\Delta\theta \rangle$, where \(\Delta\theta\) is the angle between the end-to-end vectors of a pair of particles, and angle brackets denote an average over multiple frames and trials. 
We say a spinner pair is favored when \(Q \leq -0.5\), and a mover is favored when \(Q \geq 0.5\), with coexistence \(-0.5 < Q < 0.5\). We find that spinner pairs are stable for \(\phi_{b} \leq 0.40\), while mover pairs are stable at \(\phi_{b} \geq 0.65\) [Fig. \ref{Pairing}(b)]. The \textit{lifetime} of a given type of particle pair is defined as the mean duration for which that type of pair remains together. The lifetime of a spinner pair decreases as \(\phi_{b}\) increases, dropping below $1$ s when \(\phi_b > 0.68\). In contrast, mover pairs exhibit remarkable stability, lasting over \(50\) s at \(\phi_b = 0.74\) [Fig. \ref{Pairing}(c)]. Beyond \(\phi_b =0.74\), the particle is unable to propel itself through the dense bead medium. Pairing preference and pair lifespan are closely related. 

\begin{figure}
    \centering
    \includegraphics[width=\linewidth]{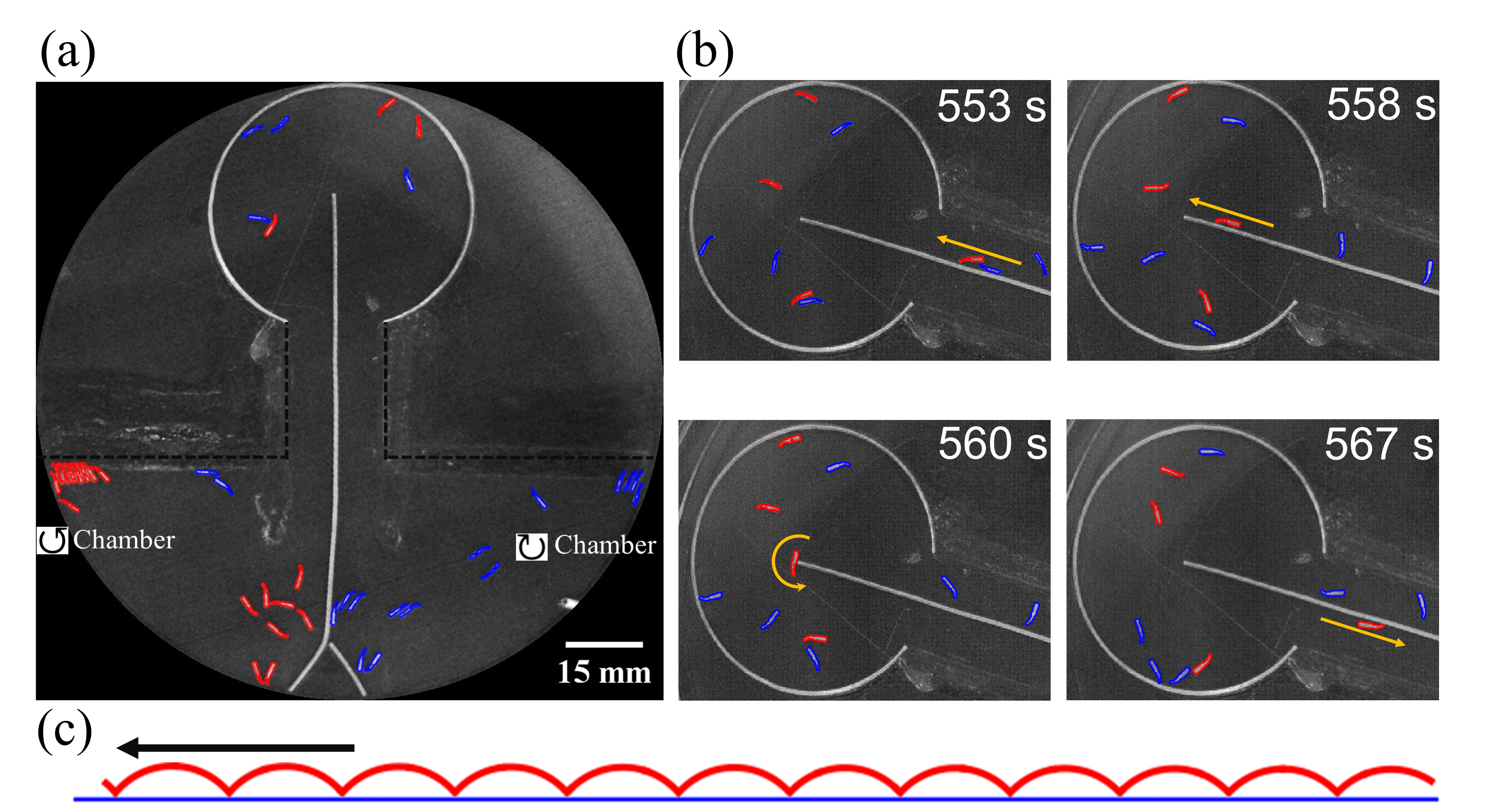}
    \caption{
    (a) Cup-straw geometry to sort motile chiral mixtures. 
    \(\boldsymbol{\circlearrowleft}\) and 
    \(\boldsymbol{\circlearrowright}\) particles from reservoir at top are separated by selectively hugging side of rail compatible with chirality of their skipping orbit.
    (b) \(\boldsymbol{\circlearrowleft}\) particles auto-correct using the same rail to go from \(\boldsymbol{\circlearrowright}\) to \(\boldsymbol{\circlearrowleft}\) chamber. (c) Section of skipping trajectory at flat wall. All parameters are as in Fig. \ref{OnlyChiral}b.} 
    \label{sorting}
\end{figure}
Since our discussion of skipping orbits shows that \( \boldsymbol{\circlearrowleft}\)/\( \boldsymbol{\circlearrowright}\) CAPPs typically have a CW/CCW orbit at a circular wall or move
keeping a straight one on its left or right, it suggests a natural mechanism for sorting CAPPs by their chirality that is significantly simpler than other methods based on using structured obstacle arrays \cite{li2022chiral,chan2024chiral} or patterned channels \cite{mijalkov2013sorting}. Each edge of a linear strip inserted into a reservoir containing a heterochiral mixture preferentially draws CAPPs of only one sign, which skate to the far end where a branch ensures their delivery into separate chambers [see Fig. \ref{sorting}  and movie SM6 \cite{moviechiralch52025}]. 
Barois et al. \cite{barois2020} also demonstrated CAPP sorting via polarized wall currents, but argues that interparticle collisions are required to initiate the transition from bulk to edge. The noisy motion of our CAPPs allows them to discover the boundary purely by diffusion and sorting occurs consistently even when a single particle is in the reservoir [movie SM7 \cite{moviechiralch52025}]. 
Defining sorting efficiency as \(N_C/(N_C+N_W)\), where \(N_C\) and \(N_W\) are the numbers of particles in the correct and wrong compartments, and starting with particle numbers ranging from 20 to 100, we consistently find an efficiency $\approx 0.9$, irrespective of the initial particle number at these densities. At higher initial densities, CAPPs tend to form dimers and do not leave the reservoir.

\section{Summary} \label{summary}
To summarize, this study presents a comprehensive experimental and theoretical investigation of the individual and pair dynamics of chiral active polar particles realised as motile vibrated grains. We have identified a robust ``skipping-orbits'' mode in which particles hug the boundary and circulate opposite to their rotation in bulk. The result is a pronounced edge accumulation, far stronger than that for achiral polar particles. Our minimal theory shows that this edge current is a natural consequence of chiral activity.
Building on these theoretical insights, we predict specific pairing behaviors in both homochiral and heterochiral particle configurations, for which we find unambiguous experimental support. Our observation and theoretical analysis of edge currents whose direction depends on chirality also leads to a simple cup and straw device for chiral sorting with high fidelity, which operates even when the sorter contains just one particle. 
Importantly, while some of these behaviours are also observed in apolar chiral particles \cite{kalz2022collisions}, the mechanism that leads to the skipping orbits crucially involves polarity. Our results demonstrate that tuning chirality provides a powerful handle for steering transport and separation without the need for elaborate micro-fabrication and complex geometrical setups.

\begin{acknowledgments} 
A.M. acknowledges the
support of ANR through the grant PSAM, and a TALENT fellowship awarded by CY Cergy Paris Universit\'e. RK thanks the UGC, India and project CRG/2023/006073 for support. SR acknowledges support from the ANRF, India, through a J C Bose Fellowship till Nov 2025 and a National Science Chair Professorship thereafter, and project CRG/2023/006073. AKS acknowledges a National Science Chair Professorship of the ANRF, Government of India. This research was supported in part by grant NSF PHY-2309135 to the Kavli Institute for Theoretical Physics.
\end{acknowledgments}

\appendix
\section{Minimality and generality of the theoretical model} 
\label{min_theory}
In this article, we have chosen to use a minimal model of our chiral active particle given by \eqref{rEq} and \eqref{pEq}.
A faithful mechanical dynamical description of a CAPP would have multiple other contributions. These include i. a self-propulsion velocity that is not along the geometric polar asymmetry axis but at an angle to it; this would enter through a chiral term $\propto\bm{\epsilon}\cdot\hat{{\bf p}}$ in \eqref{rEq}; ii. a chiral and active force density that appears as an ``odd'' mobility dotted with the force: $\mu_c\bm{\epsilon}\cdot{\bf F}$; iii. an equilibrium off-diagonal mobility that would connect velocity and torque, and angular velocity and force. We ignore these since we are primarily interested in modelling the dynamics of a single particle, and these additional terms do not lead to any qualitatively new feature not already generated by the terms we retain.


\section{The physics of the $\gamma$ coupling} 
\label{physgamma}
Our description differs from the more standard descriptions of chiral active Brownian particles \cite{Teeffelen2009Clockwise} due to the inclusion of the term $\propto\gamma$. The presence of this term in the equation of motion for $\hbf{p}$ can be attributed to more than one mechanical process. For example, $\hbf{p}$ can interact with an external potential $U$ through a term $\hbf{p} \cdot \partial_{\bf r} U$ in the energy function, which says that rotating a polar rod on a slope changes its energy, by moving portions of it into higher or lower reaches of the slope. Instead, a nonuniform drag coefficient along the rod---an extreme case of which would be the pinning of the tip by wall friction---can cause the same force to move different parts of it at different speeds and thus to rotate it. The former is a diagonal and the latter an off-diagonal kinetic coupling. Thus, although \cite{Teeffelen2009Clockwise} do not explicitly introduce the $\gamma$ term in their orientational dynamics, we expect equivalent physical effects to be present in their model.
{Indeed, the functional role of $\gamma$ does arise in our experiment primarily due to the proximity of a wall.}

More generally, a generic passive dynamics of coupled position and polarisation degrees of freedom---the latter being a unit vector---has the form
\begin{equation}
\label{reqEq}
\dot{{\bf r}}=-\mu\partial_{{\bf r}}U-{\gamma}\bm{\Pi}\cdot\partial_{{\hat{{\bf p}}}}U+\bm{\xi}_{{\bf r}}\,,
\end{equation}
\begin{equation}
\label{peqEq}
\dot{\hat{\bf p}}_i=-\bm{\Pi}\cdot\left(\Gamma_{\hat p}\partial_{{\hat{{\bf p}}}}U+{\gamma}\partial_{{\bf r}}U\right)+\bm{\xi}_{\hat{\bf p}}\,,
\end{equation}
where $\bm{\Pi}$ is defined below \eqref{pEq}, $\langle {\xi}_{{\hat{\bf p}}_i}(t) {\xi}_{{\hat{\bf p}}_j}(t')\rangle=2T\Pi_{ij}\Gamma_{\hat p} \delta(t-t')$, $\langle {\xi}_{{\hat{\bf r}}_i}(t) {\xi}_{{\hat{\bf p}}_j}(t')\rangle=\langle {\xi}_{{\hat{\bf p}}_i}(t) {\xi}_{{\hat{\bf r}}_j}(t')\rangle=2T\Pi_{ij}\gamma\delta(t-t')$, and $\langle {\xi}_{{\hat{\bf r}}_i}(t) {\xi}_{{\hat{\bf r}}_j}(t')\rangle=2T\delta_{ij}\mu\delta(t-t')$. When $\partial_{\hat{{\bf p}}}U=0$ or $\propto\hat{{\bf p}}$, this coupling does not affect the positional dynamics. However, this is not generic: for elongated particles, orientational interactions between polarities of particles $i$ and $j$ of the form $\hat{{\bf p}}_i\cdot\hat{{\bf p}}_j$ are generated by mechanics; also
see \cite{AM_asym} for an example of a microscopic interaction potential that explicitly couples polarisation with the vectorial distance between two particles (potentials of a similar form have been considered in studies of liquid crystals \cite{Selinger}). 
If $\hat{{\bf p}}_i$ is unrelated to the shape of the particle or if we are only interested in the physics of a single polar particle, as in this work, the $\partial_{\hat{{\bf p}}}U=0$ limit applies. 

The $\gamma$ term can also be viewed as arising from a \emph{reactive} coupling between \emph{velocity} and polarisation. To show this, we start with a generic passive dynamics that retains inertia for translational motion:
\begin{equation}
\label{veqEq}
m\dot{{\bf v}}+\Gamma{\bf v}=-\partial_{{\bf r}}U-\tilde{\gamma}\bm{\Pi}\cdot\partial_{{\hat{{\bf p}}}}U+\bm{\xi}_{{\bf v}}\,,
\end{equation}
\begin{equation}
\label{peqEq2}
\dot{\hat{\bf p}}=-\bm{\Pi}\cdot\left(\tilde{\Gamma}_{\hat p} \partial_{{\hat{{\bf p}}}} U-\tilde{\gamma}{\bf v}\right)+\tilde{\bm{\xi}}_{\hat{\bf p}}\,,
\end{equation}
where $\langle \tilde{{\xi}}_{{\hat{\bf p}}_i}(t) \tilde{{\xi}}_{{\hat{\bf p}}_j}(t')\rangle=2T\Pi_{ij}\tilde{\Gamma}_{\hat p} \delta(t-t')$, $\langle {\xi}_{{{\bf v}}_i}(t) {\xi}_{{{\bf v}}_j}(t')\rangle=2T\delta_{ij}\Gamma\delta(t-t')$, and $\bm{\xi}_{{\bf v}}$ and $\tilde{\bm{\xi}}_{\hat{\bf p}}$ are uncorrelated, because of the reactive nature of the coupling $\tilde{\gamma}$. We can go from \eqref{veqEq} and \eqref{peqEq2} to \eqref{reqEq} and \eqref{peqEq} by taking $m\to 0$, writing ${\bf v}$ as $\dot{{\bf r}}$, replacing ${\bf v}$ by the R.H.S. of \eqref{veqEq} divided by $\Gamma$ and identifying $\mu=1/\Gamma$, $\Gamma_{\hat p}=\tilde{\Gamma}_{\hat p}-\tilde{\gamma}^2/\Gamma$, $\gamma=\tilde{\gamma}/\Gamma$, $\bm{\xi}_{{\bf r}}=\bm{\xi}_{{\bf v}}/\Gamma$ and ${\bm{\xi}}_{\hat{\bf p}}=\tilde{{\bm{\xi}}}_{\hat{\bf p}}+(\tilde{\gamma}/\Gamma)\bm{\Pi}\cdot\bm{\xi}_{{\bf v}}$. {While ${\bf v}$ here is the velocity of a single particle, if we consider a single polar particle in a continuum fluid medium (that is frictionally screened) \cite{Brotto} or the coupled dynamics of a polarisation field and a velocity field \cite{kumar2014flocking, AM_pol}, analogous reactive couplings arise. In the former case, ${\bf v}$ should be viewed as the value of the velocity field at the site of the polar particle. In such cases, eliminating the velocity field again leads to a local coupling of the polarisation (or the polarisation field) to the force densities (unless the medium is incompressible, in which case the coupling becomes non-local \cite{Brotto, AM_pol}).}

The $\tilde{\gamma}$ coupling term emerges in a wide range of settings \cite{Sano_birds, Brotto, kumar2014flocking, Dadhichi_Jstat}. Indeed, this was discussed in some detail in a recent review \cite{self_align_rev} on ``self-aligning active particles'', in which, starting with a description that retains rotational inertia, this term was presented as a \emph{dissipative} cross-coupling between velocity and angular velocity. In this formulation, the polarisation-velocity coupling can be seen to arise from a generalised mobility (or resistance) matrix often discussed in the context of microhydrodynamics that relates forces and torques to velocities and angular velocities \cite{mihydro1, microhydro2, microhydro3}.

\begin{figure}
    \centering
    \includegraphics[width=0.5\linewidth]{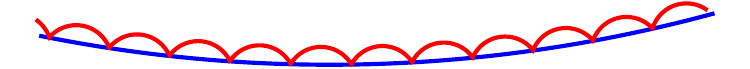}
    \caption{A section of a skipping trajectory obtained in the $\alpha=0$ case. Here, $\beta=1$, but the trajectory remains unchanged for any non-zero $\beta$ since $\zeta$ is infinite.} 
    \vspace{-0.2in}
    \label{nomob}
\end{figure}
\section{Steady state of achiral particles and measurement of $\zeta$}
\label{zeta_measure}
We start with the $\dot{r}$ and $\dot{\psi}$ equations implied by \eqref{rEq} and \eqref{pEq} with $\Omega=0$, without rescaling: 
\begin{equation}
\label{rdot1}
    \dot{r}=\mu F(r)+v_0\cos\psi
    \,,
\end{equation}
\begin{equation}
\label{psidot1}
    \dot{\psi}=-\left[F(r)\gamma+\frac{v_0}{r}\right]\sin\psi
    \,.
\end{equation}
This pair of equations have a static solution $(R_0,\psi_0)$ 
\begin{equation}
\label{SScond1}  
    {\sin\psi_0 = {R \gamma \over \mu} \sin \psi_0\cos\psi_0,\,\,\, \mu F(R_0)=-v_0\cos\psi_0.}
\end{equation}
This implies that a solution $\psi_0=0$ is stable for ${R \gamma /\mu}<1$ and another solution $\psi_0=\pm\cos^{-1}\mu/\gamma R$ exists and is stable for ${R \gamma /\mu}>1$. When $\psi_0=0$, the particle has zero velocity. In contrast, for $\psi_0=\pm\cos^{-1}\mu/\gamma R$ the orientation has an azimuthal component and the particle walks CCW or CW along the circular boundary. The existence of a nonzero solution for $\psi_0$ is a consequence of the polarity of the particle expressed through the parameter $\gamma$; such a solution does not exist for particles with $\gamma = 0$. Since $\gamma R/\mu=\zeta\rho_0$, the solutions and the conditions under which they are applicable can be written, in terms of the nondimensional quantities defined in the main text, as $\psi_0=0$ for $\rho_0\zeta<1$ or $\psi_0=\pm\cos^{-1}1/\rho_0\zeta$ for $\rho_0\zeta>1$. Note: $\rho_0$ and $\zeta$ can only be defined in chiral systems with a non-zero $\Omega$, the combination $\rho_0\zeta$ appearing here is $\Omega$-independent.

The steady-state angle for APPs can be used to estimate the ratio $\mu/\gamma$ since $R$ is the radius of our chamber. We find that APPs are disposed essentially parallel to the wall with $\psi_0\approx\pi/2$, implying that $\mu/\gamma\to 0$. 

\section{Linear stability of the steady state of chiral particles at a circular wall} 
\label{lin_stab}
We consider the stability of the static solutions $(\rho_0,\psi_0)$ in \eqref{omgcond} for $\omega=+1$. We construct the linearised equations for fluctuations $(\delta\rho$, $\delta\psi)$ about this state:
\begin{equation}
\label{fluccurv}
    \frac{d}{d\tau}\begin{pmatrix}\delta\rho\\\delta\psi\end{pmatrix}=\begin{pmatrix}\alpha F'(\bar{x}_0) &-\sin\theta_0\\-\beta\sin\theta_0F'(\bar{x}_0)&\zeta\cos^2\theta_0\end{pmatrix}\begin{pmatrix}\delta\rho\\\delta\psi\end{pmatrix}\,.
\end{equation}
The trace and determinant of the coefficient matrix in \eqref{fluccurv} are $\zeta\cos^2\psi_0+\alpha F'(\rho_0)-\cos\psi_0/\rho_0$ and $\beta\cos 2\psi_0 F'(\rho_0)-\alpha \cos\psi_0F'(\rho_0)/\rho_0+\sin^2\psi_0/\rho_0^2$, respectively. Since the former is negative and the latter positive when both eigenvalues are stabilising and as $F'(\rho_0)<0$, and $\cos2\psi_0<0$ only if $\psi_0<-\pi/4$ (recall that $0>\psi_0>-\cos^{-1}1/\zeta\rho_0$), this implies that only the more negative value of $\psi_0$ can be stable at least for large $\rho_0$ and $\zeta\equiv\beta/\alpha$.

\section{Equations of motion for a single particle at a straight wall}
\label{EOM_straight}
While in the main text we discussed the equations of motion of a single particle at a circular boundary, here we discuss the dynamics of a CAPP at a wall disposed along the $y$-axis. As for the circular boundary, we write the equations of motion in a more general form, considering a force $F_x(x)\hat{{\bf x}}$. We further define the non-dimensional variables $(\bar{x},\bar{y})\equiv (x,y)|\Omega|/v_0$. In terms of these, the equations of motion \eqref{rEq} and \eqref{pEq} become
\begin{equation}
\frac{d\bar{x}}{d\tau}=\cos\theta+\alpha F_x(\bar{x})\,,
\end{equation}
\begin{equation}
\frac{d\bar{y}}{d\tau}=\sin\theta\,,
\end{equation}
\begin{equation}
    \frac{d\theta}{d\tau}=\omega-\beta F_x(\bar{x})\sin\theta\,,
\end{equation}
where $\tau$, $\omega$, $\alpha$ and $\beta$ are defined in the main text. In a chiral system, this system of equations can lead to a state with a constant $\theta_0$, moving at a constant speed along $\hat{{\bf y}}$, with the direction determined by chirality. The solution $(\bar{x}_0,\theta_0)$ is given by
\begin{equation}
    \omega=-\zeta\sin\theta_0\cos\theta_0,\,\,\, \alpha F_x(\bar{x}_0)=-\cos\theta_0\,,
\end{equation}
which, for $|\omega/\zeta|\leq 1/2$ has two solutions between $-\pi/2$ and $\pi/2$. For $\omega=+1$, $\theta_0<0$ with one solution being between $0>\theta_0>-\pi/4$ and the other $-\pi/4>\theta_0>-\pi/2$. The linearised equations for small perturbations $(\delta\bar{x},\delta\theta)$ about these states read
\begin{equation}
\label{flucstraight}
    \frac{d}{d\tau}\begin{pmatrix}\delta\bar{x}\\\delta\theta\end{pmatrix}=\begin{pmatrix}\alpha F'(\bar{x}_0) &-\sin\theta_0\\-\beta\sin\theta_0F'(\bar{x}_0)&\zeta\cos^2\theta_0\end{pmatrix}\begin{pmatrix}\delta\bar{x}\\\delta\theta\end{pmatrix}\,.
\end{equation}
The trace and determinant of the coefficient matrix in \eqref{flucstraight} are $\zeta\cos^2\psi_0+\alpha F'(\bar{x}_0)$ and $\beta\cos 2\psi_0 F'(\bar{x}_0)$, respectively; the former has to be negative and the latter positive for stability. Since $F'(\bar{x}_0)<0$, this implies that only the solution with $-\pi/4>\psi_0>-\pi/2$ can be stable.

A CAPP at a straight wall also describes a skipping orbit. The mechanism for this is analogous to the circular case: again, $\alpha$ is immaterial; the wall force rotates the orientation of the particle; the motility, playing the role of inertia, takes the particle away from the  wall and then chirality again rotates the particle towards the wall, bringing it in contact with the wall again.

\bibliography{library}

\end{document}